\documentclass[]{spie}  

 \usepackage{float}
\usepackage{amsmath,amsfonts,amssymb}
\usepackage{graphicx}
\usepackage[colorlinks=true, allcolors=blue]{hyperref}

\title{Enhanced wavefront sensing for the Roman Coronagraph Instrument: Gaussian probes and compact model validation
}

\author[a]{Lukas Delaye}
\author[b]{Iva Laginja}
\author[a]{Pierre Baudoz}
\author[a]{Axel Potier}
\author[a]{Johan Mazoyer}
\author[a]{Raphaël Galicher}
\author[c]{Susan F. Redmond}
\author[d]{Alexis Lau}
\author[c]{A. J. Riggs}
\author[e]{Dan Sirbu}
\author[f]{Emiel H. Por}
\author[g]{Rémi Soummer}
\author[g]{Laurent Pueyo}
\affil[a]{LIRA, Observatoire de Paris, Université PSL, Sorbonne Université, Université Paris Cité, CY Cergy Paris Université, CNRS, 92195 Meudon, France}
\affil[b]{Université Côte d'Azur, Observatoire de la Côte d'Azur, CNRS, Laboratoire Lagrange, Bd de l'Observatoire, CS 34229, 06304 Nice cedex 4, France}
\affil[c]{Jet Propulsion Laboratory, California Institute of Technology, 4800 Oak Grove Dr, Pasadena, CA 91011, USA}
\affil[d]{Aix Marseille Univ, CNRS, CNES, LAM, Marseille, France}
\affil[e]{NASA Ames Research Center, Moffett field, CA 94035, USA}
\affil[f]{Department of Astronomy \& Astrophysics, University of California, Santa Cruz, CA 95064, USA}
\affil[g]{Space Telescope Science Institute, 3700 San Martin Dr., Baltimore, MD 21218, USA}
\authorinfo{Send correspondence to lukas.delaye@obspm.fr}
\begin{document} 
\maketitle

\begin{abstract}
The Coronagraph Instrument on the Roman Space Telescope will be the first space-based system to demonstrate closed-loop focal-plane wavefront sensing and control, a key step towards the Habitable Worlds Observatory. Beyond the baseline Hybrid Lyot Coronagraph, ``enhanced modes'' are being developed to improve efficiency and science yield. One such mode uses Gaussian probes for electric field estimation, extending the linear regime and allowing higher probe amplitudes. This may increase signal-to-noise, reduce exposure time, accelerate dark hole convergence, and extend operation to stars as faint as $V \sim 5$. For those reasons, it was selected by the Coronagraph Community Participation Program’s Hardware Working Group as the first technology demonstration carried out on Roman in early 2027. We present numerical simulations using a noise-free compact software model, which demonstrate the benefits of replacing the nominal probes with Gaussian probes.
\end{abstract}


\section{INTRODUCTION: the Roman Coronagraph Instrument}
\label{sec:intro}  
 The Roman Space Telescope (RST)~\cite{poberezhskiy2021roman} is a space observatory, scheduled for launch on August 30, 2026, featuring a primary mirror with a geometric diameter of 2.4 meters. Onboard, one of its instruments is the Coronagraph Instrument~\cite{creager2025coronagraph}, which serves as a technology demonstrator in preparation for the Habitable Worlds Observatory (HWO)~\cite{coyle2024space}. The primary goal of RST is to achieve unprecedented levels of contrast for direct imaging of exoplanets, using a coronagraph to block out the light from the host star.

\begin{figure}[ht]
  \begin{center}
    \begin{tabular}{c} 
      \includegraphics[trim= 4.1cm 20.5cm 0cm 2cm, clip, height=7cm]{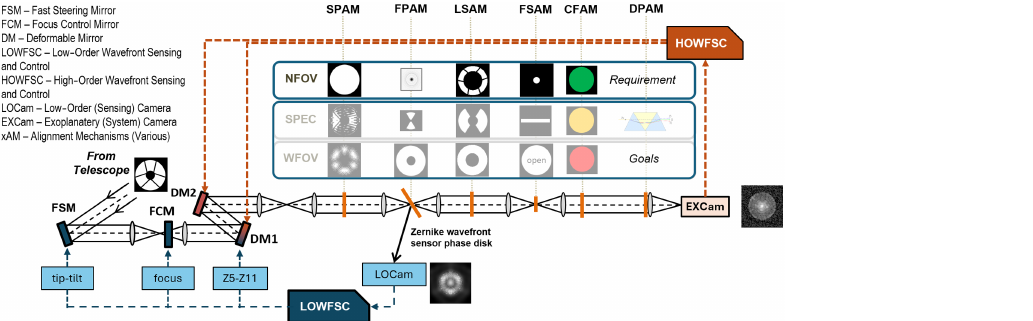}
    \end{tabular}
\end{center}
  \caption[example] 
  { \label{fig:example} 
  Unfolded Roman Coronagraph Instrument optical layout~\cite{creager2025coronagraph}. The part simulated by \texttt{corgihowfsc} is the high-frequency control loop \textsc{HOWFSC} (high-order wavefront sensing and control).}
\end{figure}

The Roman Coronagraph Instrument has several modes and spectral bands, which are shown in Fig.~\ref{fig:example} and detailed in Table~\ref{tab:cgi_modes_spectraux}. The study presented here was conducted in Band 1 narrow field-of-view (NFOV) imaging. Operating around 575 nm with a bandwidth of 10\%, this mode uses the Hybrid-Lyot Coronagraph (HLC) architecture to maximize detection sensitivity in an area of 3 to 9 $\lambda/D$. This band is one of the preferred bands for the direct detection of exoplanets. Alongside this primary mode, the instrument also includes a Wide Field-Of-View (WFOV) imaging mode and spectroscopic modes that rely on a Shaped Pupil Coronagraph (SPC) architecture.

\begin{table}[H]
    \centering
    \renewcommand{\arraystretch}{1.3} 
    \caption{Wavelength bands observed by Roman Coronagraph Instrument (adapted from~\cite{cady_high-order_2025, riggs_flight_2025, krist2023end}).}
    \label{tab:cgi_modes_spectraux}
    \begin{tabular}{l c c c c l}
        \hline\hline
        \textbf{Band} & \textbf{Mode} & \textbf{$\lambda_c$ (nm)} & \textbf{$\Delta\lambda/\lambda$, Width (nm)} & \textbf{Coronagraph} & \textbf{DH size} \\
        \hline
        Band 1 & NFOV Imaging & 575 & 10\,\%, [546, 604] & HLC & 3 to 9 $\lambda/D$ ($360^\circ$) \\
        Band 2 & Spectroscopy & 660 & 15\,\%, [610, 710] & SPC & 3 to 9 $\lambda/D$ (Bowtie) \\
        Band 3 & Spectroscopy & 730 & 15\,\%, [675, 785] & SPC & 3 to 9 $\lambda/D$ (Bowtie) \\
        Band 4 & WFOV Imaging & 825 & 10\,\%, [783, 867] & SPC & 6 to 20 $\lambda/D$ ($360^\circ$) \\
        \hline\hline
    \end{tabular}
    
\end{table}

However, a coronagraph alone is not sufficient. Wavefront aberrations propagating through the telescope optics inevitably cause starlight to leak through the coronagraphic masks, creating residual stellar light patterns, known as speckles, on the science camera. To minimize these speckle intensities and achieve the required sensitivity, the telescope features a high-order wavefront sensing and control (HOWFSC) system coupled with two active deformable mirrors (DMs)~\cite{cady_high-order_2025} (see Fig.~\ref{fig:example}). The primary objective of this active control loop is to dig a ``dark hole'' (DH), a targeted high-contrast area in the scientific detector area. 

Within this architecture, the residual speckle electric field must first be measured. This is achieved through focal-plane wavefront sensing via a temporal modulation of the speckles, performed by applying well-known voltages to the DMs called probes. This method is named Pairwise Probing (PWP)~\cite{giveon_closed_2007}. Once the field is estimated, the electric field conjugation (EFC)~\cite{giveon_closed_2007} algorithm is used to calculate the new voltages to be applied to the DMs. This minimizes the intensity of the speckles in the focal plane~\cite{giveon_broadband_2007}. It is within this specific focal-plane wavefront sensing step (PWP) that our study is situated.

Due to the strict computational and memory limitations of the spacecraft's onboard flight computer, the Roman Coronagraph Instrument operates this HOWFSC loop through a Ground-In-The-Loop (GITL) operational concept~\cite{poberezhskiy2021roman}. Instead of computing the complex control matrices onboard, the computational part is shifted to the ground. The PWP estimation and EFC inversion are part of the calculations performed on Earth before sending commands back to the spacecraft.

The operational performance of this HOWFSC loop will be evaluated on-sky during the first Technology Demonstration (TD1) campaign of the Roman Coronagraph Instrument~\cite{kasdin2020nancy}, scheduled for early 2027. Currently, the baseline PWP algorithm relies on combinations of Sinc-sinc-sine probes to dig the dark hole~\cite{cady_high-order_2025}. However, to maintain a sufficient signal-to-noise ratio on faint targets, the amplitude of these probes must be increased, which inherently breaks the linear approximation of the electric field and degrades the estimation with nonlinear biases. The recent study by Laginja et al. (2025)~\cite{laginja_extended_2025}, has shown that highly localized probes, such as a single poked actuator, are significantly more resilient to these nonlinearities than the nominal probes. Building on this concept, we investigate the use of Gaussian probes that are better adapted to the optical design of the Roman Coronagraph Instrument. They offer the same linear behavior as single actuators, but provide smoother energy modulation across the targeted focal plane area for a given amplitude.

The primary objective during the TD1 campaign for Gaussian probing is to validate the implementation and stability of these new probe profiles on the flight DMs in Band 1 and to demonstrate their resilience to the nonlinear estimator biases that typically degrade nominal probing. To do so, we rely on two distinct simulation frameworks: \texttt{corgihowfsc}\footnote{\url{https://github.com/roman-corgi/corgihowfsc}}, the official control software pipeline for the Roman Coronagraph Instrument which features a compact optical model to validate flight algorithms, and \texttt{Asterix}\footnote{\url{https://github.com/johanmazoyer/Asterix}}~\cite{mazoyer2019asterix}, a public package that allows fast closed-loop simulations and that contains the compact optical model of the THD2 testbed~\cite{baudoz2018optimization} (testbed at LIRA / Observatoire de Paris – PSL in Meudon, France) with the HLC~\cite{laginja_extended_2025}.

\section{Estimation of the electric field using PWP}
\label{sec:pairwise_probing}
PWP is an algorithm designed to estimate the electric field of the speckles in the focal plane. A set of phase probes $\pm\Psi_m$ is introduced on the DM, where the index $m$ denotes a specific probe. Given the images obtained when the probes are applied $(I^+_m, I^-_m)$ and the electric field of the probes alone, solving an inverse problem numerically allows us to estimate the electric field of the speckles $E_s$. A summary of the algorithm is provided in Appendix~\ref{annexe}.
\subsection{Analytical Study of PWP Nonlinearities}\label{NL_analytic}

 To achieve a high signal-to-noise ratio during the pairwise estimation, especially for faint targets or to reduce exposure times, it is highly desirable to increase the amplitude of the probe by increasing $\Psi_m$~\cite{groff2016methods}. However, large probe amplitudes break the linear approximation of the electric field used in Appendix~\ref{annexe_1}. Upon applying a probe, the electric field propagates through the coronagraphic operator $C$, yielding the following focal plane intensities:
\begin{align}
    I_m^+ &= |C[E_0e^{i\Psi_m}]|^2 + I_{\text{incoh}} \label{I^+}\, ,  \\
    I_m^- &=  |C[E_0e^{-i\Psi_m}]|^2 + I_{\text{incoh}} \, , 
    \label{I^-}
\end{align}
where we define the coherent stellar speckle field $E_s = C[E_0]$ and the residual incoherent light $I_{\text{incoh}}$ (e.g., exoplanet signal). Following the analytical development presented by Groff et al. (2016)~\cite{groff2016methods} and Laginja et al. (2025)~\cite{laginja_extended_2025}, we can expand the exponential operator of the probed pupil field $E_0 e^{\pm i\Psi_m}$ using a Taylor series up to the fourth order:
\begin{equation}
    E_0 e^{\pm i\Psi_m} \approx E_0 \left(1 \pm i\Psi_m - \frac{1}{2}\Psi_m^2 \mp \frac{i}{6}\Psi_m^3 + \frac{1}{24}\Psi_m^4 \right) \, .
\end{equation}

Propagating this electric field through the coronagraphic operator $C$ yields the complex electric field in the focal plane:
\begin{equation}
    C[E_0 e^{\pm i\Psi_m}] \approx E_s \pm iC[E_0\Psi_m] - \frac{1}{2}C[E_0\Psi_m^2] \mp \frac{i}{6}C[E_0\Psi_m^3] + \frac{1}{24}C[E_0\Psi_m^4] \, . \label{NL_dev}
\end{equation}

Using Eq.~\ref{I^+}, \ref{I^-} and \ref{NL_dev} to express $I_+ - I_- = 2\delta_{\mathrm{measured},m}$, and keeping all terms up to the fifth order, we obtain:
\begin{align}
    \delta_{\mathrm{measured},m} \simeq \quad & 2\Im\Big(E_s C[E_0\Psi_m]^*\Big) & (\text{Term I}) \nonumber \\ 
    + & \Im\Big(C[E_0\Psi_m] C[E_0\Psi_m^2]^*\Big) & (\text{Term II}) \nonumber \\
    - & \frac{1}{3}\Im\Big(E_s C[E_0\Psi_m^3]^*\Big) & (\text{Term III}) \nonumber \\
    + & \frac{1}{6}\Im\Big(C[E_0\Psi_m^2] C[E_0\Psi_m^3]^*\Big) & (\text{Term IV}) \nonumber \\
    - & \frac{1}{12}\Im\Big(C[E_0\Psi_m] C[E_0\Psi_m^4]^*\Big) \,  & (\text{Term V}) \label{eq:nonlinear_terms}
\end{align}

Term I is the desired linear approximation used in standard PWP (corresponding to the first-order field, identical to our derivation in Section \ref{annexe_2}). Terms II through V represent the nonlinear estimation biases that arise at high probe amplitudes.

Term II is a third-order cross-term and constitutes the dominant source of nonlinear error. Crucially, term II depends only on the probe morphology and is entirely independent of the residual speckle field $E_s$. Term III also arises from third-order effects, but because it scales with $E_s$, its impact naturally decreases as the DH deepens and the speckle intensity is minimized. Terms IV and V are fifth-order terms and remain generally negligible with respect to terms II and III.

Consequently, to extend the linearity range of the PWP algorithm and use probes with higher amplitude, the probe shape must be optimized to minimize Term II. In this paper, we show that using a Gaussian shape instead of the nominal sinc-sinc-sine shape can minimize nonlinear terms in Eq.~\ref{eq:nonlinear_terms}, thereby minimizing nonlinear residuals. 

\subsection{Theoretical Formulation of the Nonlinear PWP Error}\label{TheoryNL}
\subsubsection{Nonlinear Error in PWP Estimation}
Choosing another probe shape is justifiable if it improves estimation performance. In particular, a different type of probe can help mitigate the nonlinear effects described in Sec.~\ref{NL_analytic}. With the noise-free model (no detector and photon noise), our idea is first to compare two simulated estimations with two different probe shapes, to show that one estimation is more degraded by nonlinear effects than the other.

A useful metric for measuring these nonlinear errors is the residual of the estimated intensity using PWP relative to the speckle intensity on the detector. This choice is motivated by coherent differential imaging (CDI) techniques, initially investigated by Bottom et al. (2017)~\cite{bottom2017speckle}, Jovanovic et al. (2018)~\cite{jovanovic2018review} and Potier et al. (2022)~\cite{potier_increasing_2022}. In the CDI formalism, we use PWP to produce the estimate $I_{\mathrm{PWP}}$ of the true speckle intensity $I_s=|E_s|^2$. We then subtract this estimate from the total intensity $I_0 = I_s+I_{\mathrm{incoh}}$ on the detector and obtain the CDI image:
\begin{equation}
    I_{\mathrm{CDI}} = I_0 - I_{\mathrm{PWP}} = I_{\mathrm{incoh}} + I_{\mathrm{PWP,NL}}   \, ,
\end{equation}
where $I_{\mathrm{incoh}}$ is the incoherent intensity and $I_{\mathrm{PWP,NL}}$ is the residual of the coherent starlight due to the nonlinear PWP errors. Assuming no incoherent light (no off-axis sources, $I_{\mathrm{incoh}} =  0$):  
\begin{equation}
    I_{\mathrm{CDI}}= I_{\mathrm{PWP,NL}}  \, .
\end{equation}

The CDI intensity thus equals the nonlinear PWP errors. In this section, we express its spatial variance. 

We first separate the estimated coherent speckle intensity into its complex field components:
\begin{equation}
    I_{\mathrm{PWP}} = |E_{\mathrm{PWP}}|^2 = \Re(E_{\mathrm{PWP}})^2 + \Im(E_{\mathrm{PWP}})^2 \, .
\end{equation}

We assume that the nonlinear errors are included into the second term of Eq.~(64) in Groff et al. (2016)~\cite{groff2016methods}:
\begin{equation}
    \begin{bmatrix}
        \sigma^2_{\mathrm{NL}}(\Re(E_{\mathrm{PWP}})) \\
        \sigma^2_{\mathrm{NL}}(\Im(E_{\mathrm{PWP}}))
    \end{bmatrix}
    =
    \frac{\sigma_{P}^2}{I_{\mathrm{probe}}}
    \begin{bmatrix}
        \Re(E_{\mathrm{PWP}})^2 \\
        \Im(E_{\mathrm{PWP}})^2
    \end{bmatrix} \, ,
    \label{Matrice_deltap}
\end{equation}
where $I_{\mathrm{probe}}$ is the focal plane intensity of a single probe (assumed identical for all probes $m$), and $\sigma_P^2$ is the sum over the probe pairs of the variances of the nonlinear model error on the probe electric field. The evaluation of $\sigma_P^2$ is detailed in the following section. 

Assuming $\sigma_{\mathrm{NL}}(\Re(E_{\mathrm{PWP}}))$ and $\sigma_{\mathrm{NL}}(\Im(E_{\mathrm{PWP}}))$ are perfectly correlated, the statistical variance of~$I_{\mathrm{PWP,NL}}$ is obtained by squaring the sum of the correlated field differentials (see Appendix \ref{annexe_correlation}):
\begin{equation}
    \sigma_{I_{\mathrm{PWP,NL}}}^2 = \left( 2\Re(E_{\mathrm{PWP}})\,\sigma_{\mathrm{NL}}(\Re(E_{\mathrm{PWP}})) + 2\Im(E_{\mathrm{PWP}})\,\sigma_{\mathrm{NL}}(\Im(E_{\mathrm{PWP}})) \right)^2 \, ,
    \label{correlation_termes}
\end{equation}
which simplifies as:
\begin{equation}
    \sigma_{I_{\mathrm{PWP,NL}}}^2 = \frac{4\,\sigma_P^2}{I_{\mathrm{probe}}}\,I^2_{\mathrm{PWP}}
    \, .\label{erreur_im_re}
\end{equation}

The only unknown in Eq~\ref{erreur_im_re} is $\sigma_{P}$ which is studied in the next section.

\subsubsection{Nonlinear Error in Probe Intensity}
\label{non_linear_error_probe}
To evaluate $\sigma_{P}^2$, we first isolate the nonlinear PWP error at the modulation level. We define the nonlinear error as the residual between the measured and linear modulations:\begin{equation}
    \delta_{m,\mathrm{NL}} = \delta_{\mathrm{measured},m} - \delta_m \, ,
\label{eq:delta_nl}
\end{equation} 
where $\delta_{\mathrm{measured},m}$ corresponds to the real calculation of $I^+_m-I^-_m$ defined in Eq~\ref{eq:nonlinear_terms}, which contains nonlinear terms. The term $\delta_m$ corresponds to the linear term in the PWP algorithm (term I in Eq~\ref{eq:nonlinear_terms}). 

To link the analytical model to the simulations, we define a global fractional error ``$\mathrm{err}_m$'' for each probe $m$, evaluated spatially across the pixels of the DH. We compute this scalar factor as the ratio of the spatial standard deviation (STD) of the nonlinear error (Eq.~\ref{eq:delta_nl}) to the spatial STD of the linear modulation~$\delta_m$:
\begin{equation}
    \mathrm{err}_m = \frac{\mathrm{STD}_{\mathrm{DH}}(\delta_{m,\mathrm{NL}})}{\mathrm{STD}_{\mathrm{DH}}(\delta_m)}\, .
\end{equation}

We assume that this fractional error is approximately uniform across the focal plane, meaning that the amplitude of the local nonlinear error scales proportionally with the amplitude of the linear modulation:
\begin{equation}
    |\delta_{m,\mathrm{NL}}| \approx \mathrm{err}_m |\delta_m| \, .
\end{equation}

This local absolute error per resolution element is then used as a proxy for the local statistical uncertainty:
\begin{equation}
    \sigma_{\delta_{m,\mathrm{NL}}} \approx |\delta_{m,\mathrm{NL}}| \, .
\end{equation}

Moreover, using the linear approximation, we can write the absolute value of $\delta_m$ as follows: 
\begin{equation}
    |\delta_m| = |2\Im(E_sP_m^*)| \propto \sqrt{I_sI_{\mathrm{probe}, m}}\qquad\Rightarrow\qquad     I_{\mathrm{probe}, m} \propto \frac{\delta_m^2}{I_s}\, ,
    \label{I_probe_m}
\end{equation}
where $I_{\mathrm{probe}, m} = |P_m|^2 \equiv |C[E_0\Psi_m]|^2$ is the intensity of the probe in the focal plane. 

Therefore, a model error on $\delta_m$ (e.g., nonlinearities) induces a model error on $I_{\mathrm{probe}, m}$ and the spatial variance on the modulated intensity $\sigma^2_{I_{\mathrm{probe}, m}}$ comes from the PWP nonlinearity variance $\sigma^2_{\delta_{m,\mathrm{NL}}}$: 
\begin{equation}
   \sigma_{I_{\mathrm{probe}, m}}^2 \propto 4\frac{\delta_m^2}{I_s^2} \sigma_{\delta_{m,\mathrm{NL}}}^2 = 4 \mathrm{err}_m^2 I_{\mathrm{probe}, m}^2\, .
\end{equation} 

By propagating the intensity variance to the probe amplitudes, which are all equal ($|P_m| = \sqrt{I_{\mathrm{probe}}}$), the variance of the model error for a single probe $m$ becomes:
\begin{equation}
    \sigma_{P_m}^2 = \left( \frac{1}{2\sqrt{I_{\mathrm{probe}}}} \right)^2 \sigma_{I_{\mathrm{probe}, m}}^2 = \frac{1}{4I_{\mathrm{probe}}}\sigma_{I_{\mathrm{probe}, m}}^2 = I_\mathrm{probe}\mathrm{err}_m^2\,.
\end{equation}

In a good approximation, the total variance $\sigma_{P}^2$ can be estimated as the sum of the individual variances:
\begin{equation}
    \sigma_{P}^2 \approx  \sum_{m=1}^3 \sigma_{P_m}^2 = I_{\mathrm{probe}} \left( \sum_{m=1}^3 \mathrm{err}_m^2 \right) = I_{\mathrm{probe}}\mathrm{err}^2 \, ,
    \label{err_prop}
\end{equation}
defining an effective global nonlinear error $\mathrm{err}$ as $\mathrm{err} = \sqrt{\sum\limits_{m=1}^{3} \mathrm{err}_m^2}$.

For a noise-free simulation, Eqs.~\ref{erreur_im_re} and~\ref{err_prop} write:
\begin{equation}
        \sigma^2_{I_{\mathrm{PWP,NL}}} = 4\, I_{\mathrm{PWP}}^2 \,\mathrm{err}^2 \, .
        \label{sigma_Incoherent}
\end{equation}

 Finally, the global spatial variance measured across the CDI image ($\sigma^2_{I_{\mathrm{CDI}}}$) serves as a direct macroscopic estimator for the local statistical error. We can therefore write:
\begin{equation}
    \sigma^2_{I_{\mathrm{CDI}}} = \sigma^2_{I_{\mathrm{PWP,NL}}} = 4I_{\mathrm{PWP}}^2\mathrm{err}^2\,.
\end{equation}

It thus turns out that the spatial variance of the CDI residuals $\sigma^2_{I_{\mathrm{CDI}}}$ is directly proportional to $\mathrm{err}^2\,I_{\mathrm{PWP}}^2$.

\section{Results with Numerical simulations}\label{results}
\subsection{Comparison of Gaussian Probes with Nominal Probes using \texttt{Asterix}}
\label{sim_gaussian}
To begin with, we analyzed the shape of the Gaussian probe we put on the DM to determine its optimal parameters, using the optical compact model provided with \texttt{Asterix}. The choice of Gaussian parameters is crucial for obtaining a DH illumination that is as close as possible to that obtained with the default sinc-sinc-sine probe set. 
\subsubsection{Selecting the parameters of the Gaussian probe}
First, we want to ensure that the energy in the focal plane of the Gaussian probe is substantially identical to what is found with the use of Sinc-sinc-sine probes. In the DM plane, the Gaussian probe is fully characterized by the location $(x_m,y_m)$ of its maximum value, its amplitude $A_m$, and its standard deviation $\sigma_G$:
\begin{equation}
    \Psi_m(x, y) = A_m \exp\left( - \frac{(x - x_m)^2 + (y - y_m)^2}{2\sigma_G^2} \right) \, ,
\end{equation}
where $(x, y)$ are the spatial coordinates in the DM plane.

\begin{figure}[H]
    \centering
    \includegraphics[width=1\linewidth]{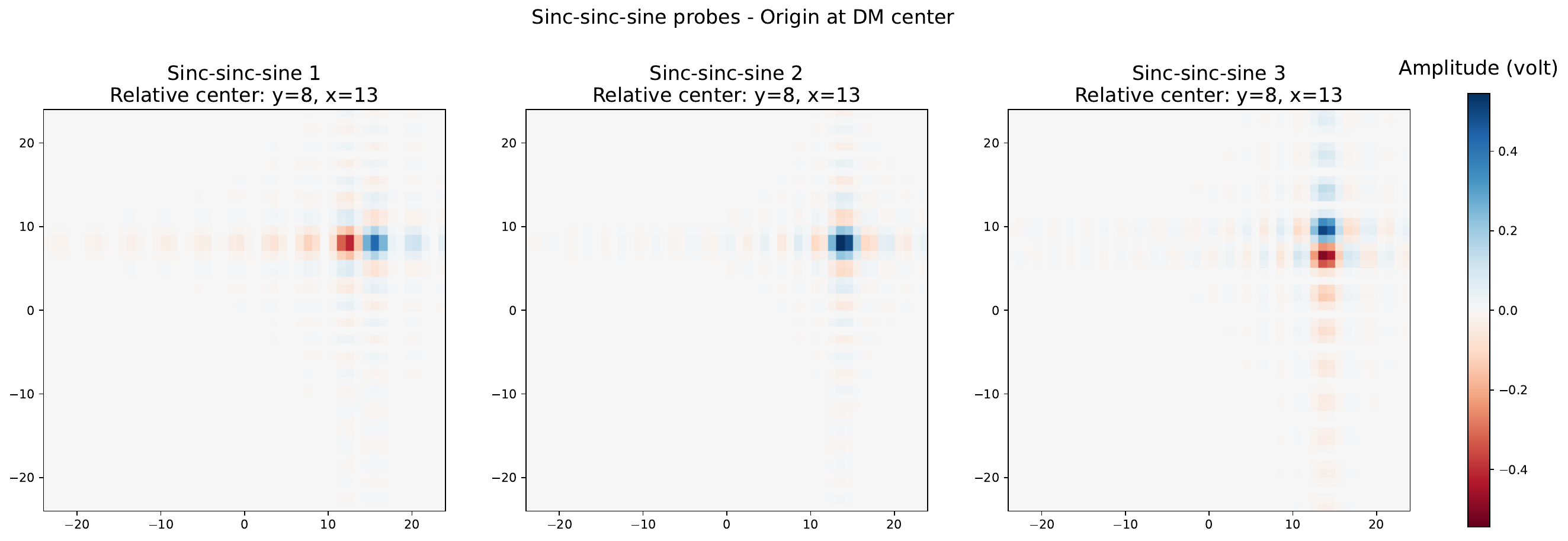}
     \includegraphics[width=1\linewidth]{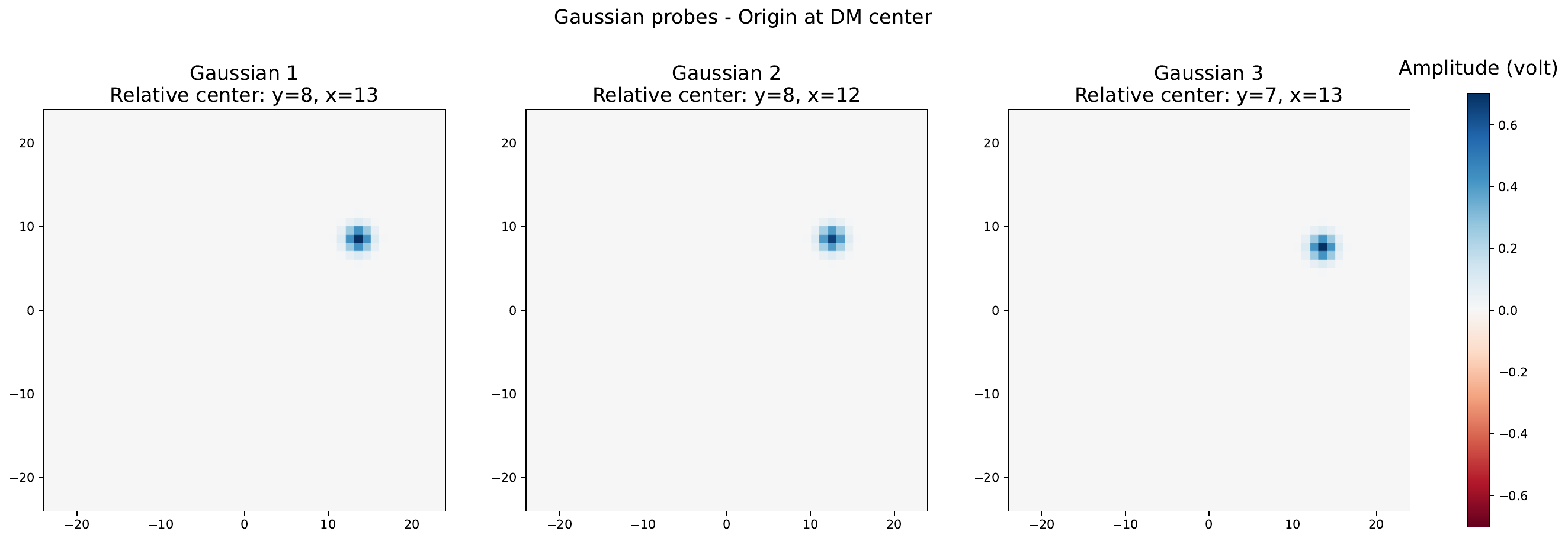}
\caption{DM actuator voltages corresponding to the nominal sinc-sinc-sine probes and the alternative Gaussian probes.}
\label{probe_shapes}
\end{figure}

As shown in Fig.~\ref{probe_shapes}, we positioned the three Gaussian probes at $(x_1=13,y_1=8)$, $(x_2=12,y_2=8)$ and $(x_3=13,y_3=7)$. We implemented this configuration to ensure that at least two probes provide distinct intensity modulations for each pixel $(k,l)$ in the focal plane.

For TD1, we potentially want to set the mean intensity in the DH to $5\times 10^{-7}$. Such a value would be a good choice for on-sky testing at $2\times10^{-8}$ contrast, as it allows for strong modulation of the speckle intensity with a good signal-to-noise ratio, potentially highlighting nonlinear effects. Finally, we must choose a value for $\sigma_G$ that best concentrates the modulation energy in the DH. The contrast on the detector before starting PWP is approximately $2\times 10^{-8}$ for these simulations.
\begin{figure}[H]
    \centering
    \includegraphics[width=0.68\linewidth]{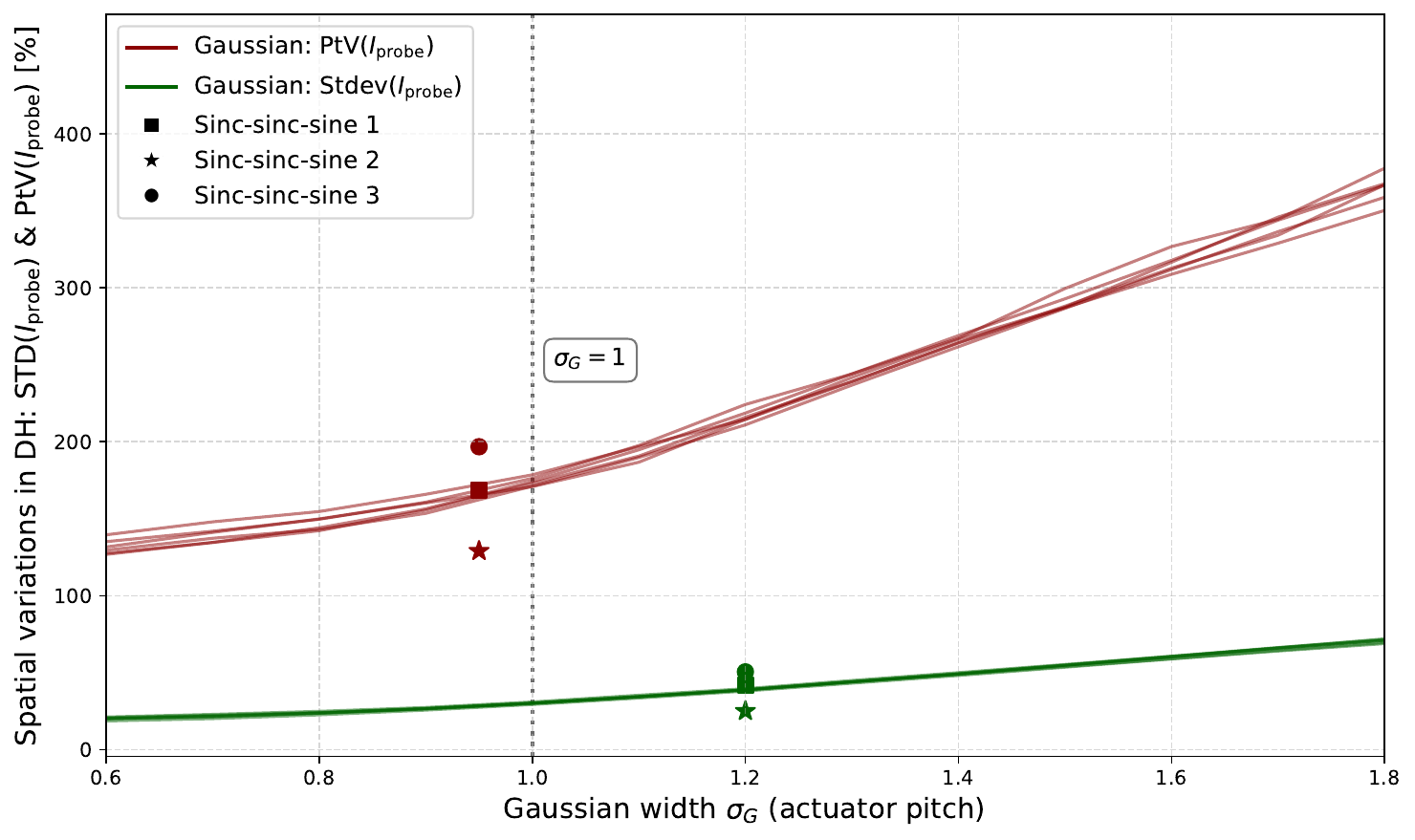}
    \caption{Spatial variation of the DH intensity as a function of the Gaussian probe size, simulated with \texttt{Asterix}. The six red lines show the peak-to-valley amplitude for each of the six individual probes, while the six green lines represent their respective spatial standard deviations. For comparison, the data points for the Sinc-sinc-sine probes are included, corresponding to the average of each probe pair.}
    \label{sigma_asterix}
\end{figure}

In Fig.~\ref{sigma_asterix}, we plot the Peak-to-Valley intensity for the sinc-sinc-sine probes, and we notice that their means correspond to the Peak-to-Valley intensity of Gaussian with $\sigma_G = 0.95$ pitch. Then, we plot the standard deviation of the intensity for the sinc-sinc-sine probes, and we notice that it corresponds to the standard deviation intensity of Gaussian with $\sigma_G = 1.2$ pitch.

We know that sinc-sinc-sine probes have been optimized to inject light solely into the DH. Since we want our Gaussian probes to have the same values as the sinc-sinc-sine probes, an intermediate value of $\sigma_G = 1$ is a good compromise for the Gaussian probes used in our study.

\subsubsection{Simulation of nonlinear errors in the PWP}
\begin{figure}[H]
    \centering
    \includegraphics[width=0.68\linewidth]{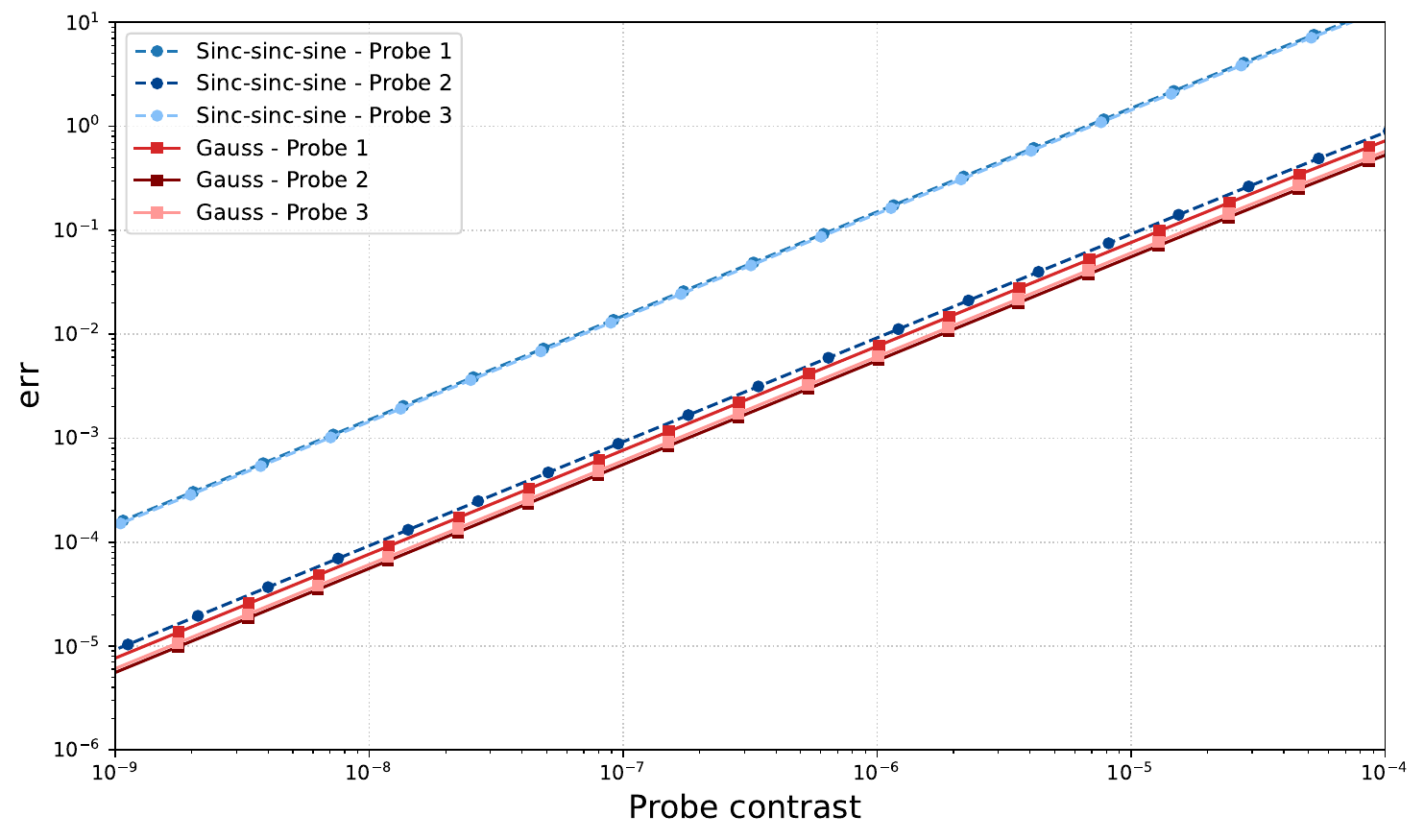}
    \caption{Numerical simulation using \texttt{Asterix} of the nonlinear error as a function of the probe contrast. The probe contrast is the average contrast in the focal plane modulated by each of the probes.}
    \label{err_NL}
\end{figure}

In Fig.~\ref{err_NL}, we estimate with \texttt{Asterix} the value of $\mathrm{err}_m$ on the probe $m$, for the three nominal sinc-sinc-sine probes and three Gaussian probes. We assume a mean speckle intensity contrast of $2\times 10^{-8}$ in the DH, and we simulate the error value as a function of the probe contrast in the DH. Note that the value of the nonlinear error is derivable only in numerical simulation, by isolating the linear and nonlinear terms of Eq.~\ref{eq:nonlinear_terms}. 

On one hand, we can see that all three Gaussian probes share a similar behavior, although slightly lower than the unmodulated sinc probe (nonlinear error of 0.46\% for the unmodulated sinc versus 0.38 \% for the highest Gaussian with a probe contrast of $5\times10^{-7}$). On the other hand, the other two modulated sinc-sinc-sine probes have much higher errors than the three Gaussian probes (nonlinear error up to 7.5\% both for the modulated sinc with a probe contrast of $5\times10^{-7}$). A linear interpolation between the error and the probe contrast was made with Fig.~\ref{err_NL}, and it shows that $\mathrm{err}_m \propto I_{\mathrm{probe,m}}$, regardless of the probe shape.
\subsection{First Simulation with \texttt{corgihowfsc} Compact model}
\label{cgi}
This section aims to validate the nonlinear error model built in Sec.~\ref{TheoryNL}. We assess the CDI performance with both probe sets in ideal condition, using the compact model of \texttt{corgihowfsc}. Notably, this compact model is noise-free, which means that we neglect all sources of errors except nonlinearities. 

The different probes to be compared are shown in Fig.~\ref{probe_shapes}. They are scaled to set a contrast of $5 \times 10^{-7}$ in the focal plane after application to the DM.

\begin{figure}[H]
    \centering
    \begin{minipage}{0.40\textwidth}
        \centering
        \includegraphics[ width=\linewidth]{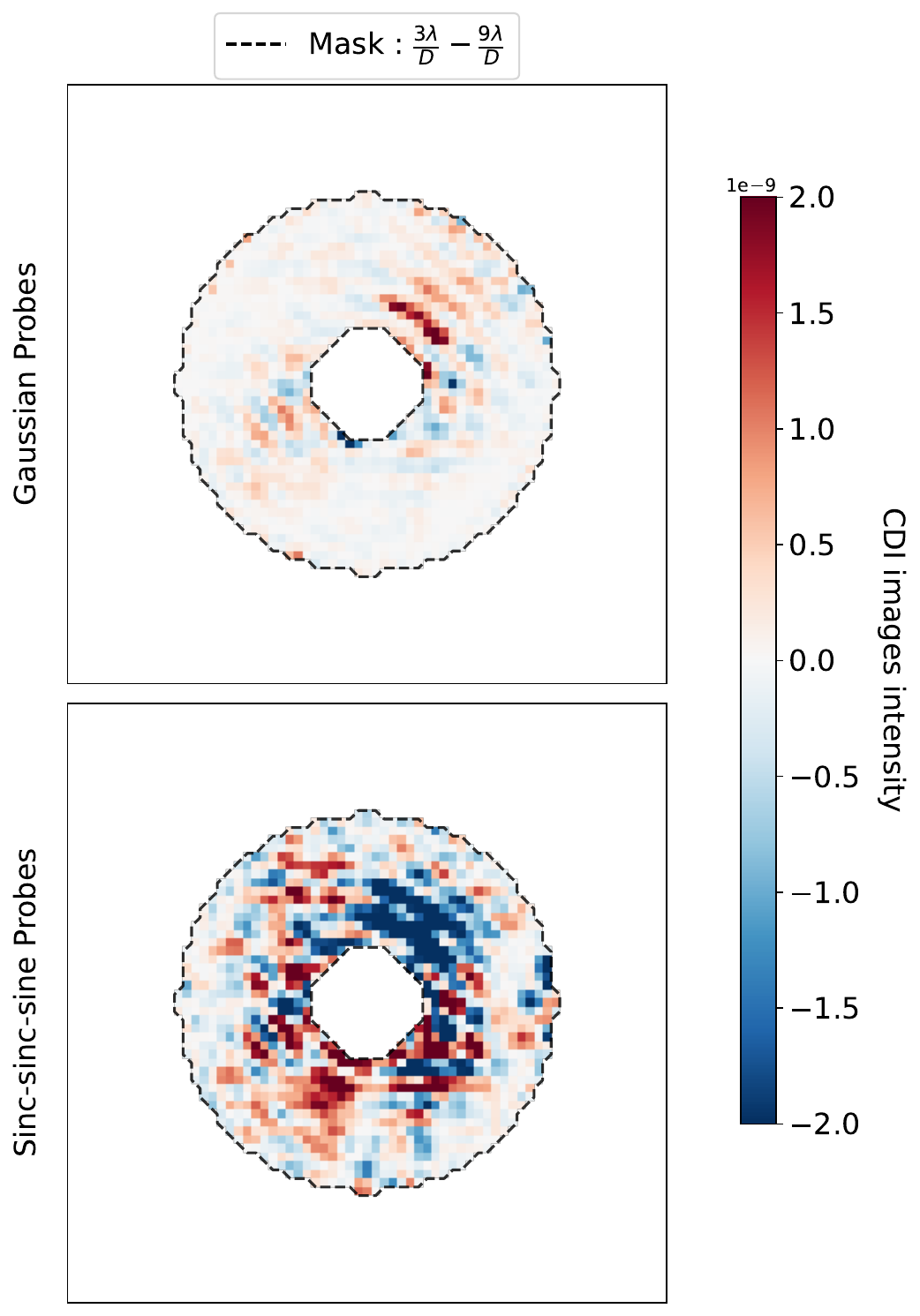}
    \end{minipage}\hfill
    \begin{minipage}{0.6\textwidth}
        \centering
        \includegraphics[trim=0cm 0cm 0cm 0cm, clip, width=\linewidth]{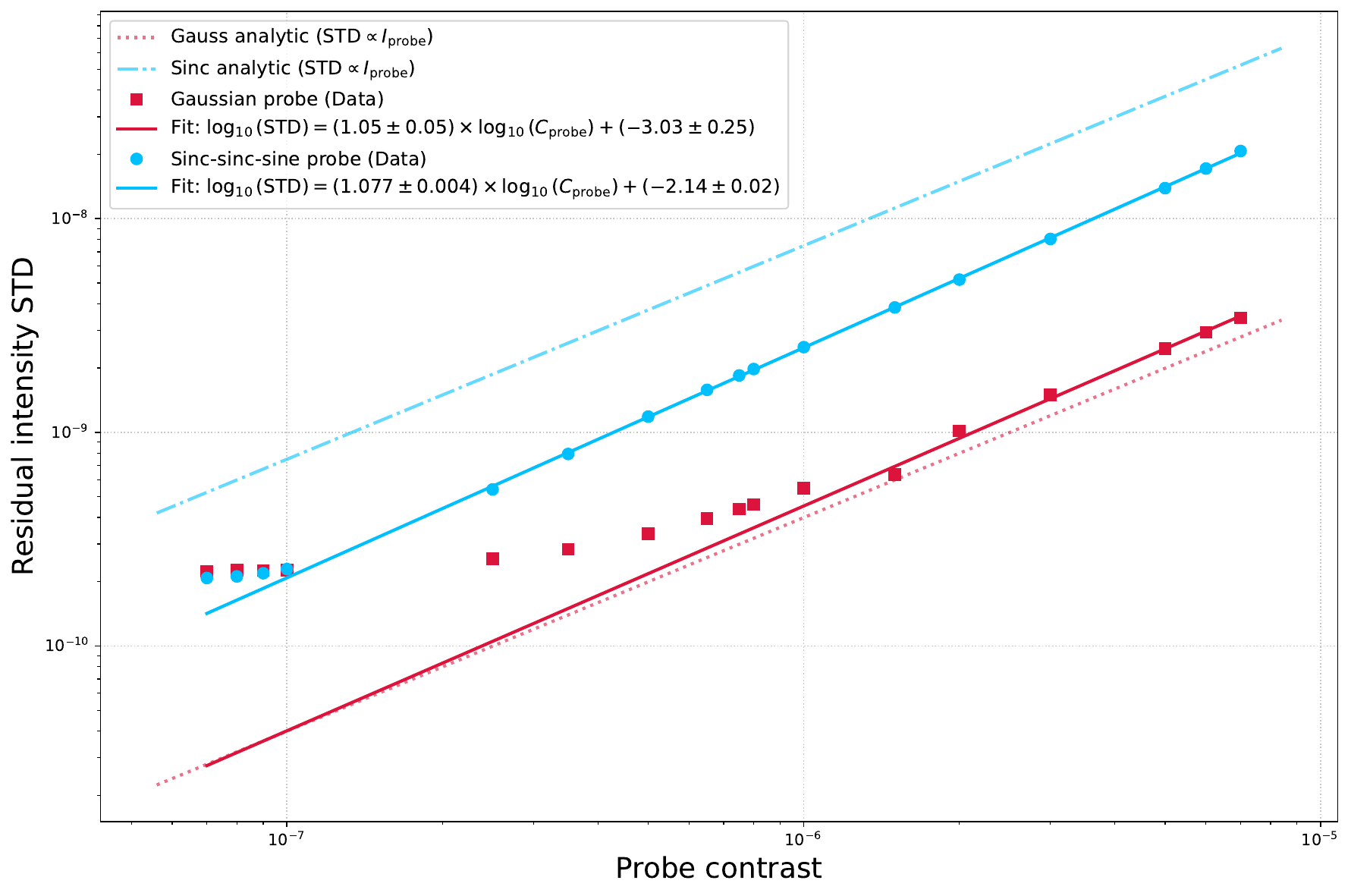}
    \end{minipage}
    \caption{\textbf{Left:} Intensity in the CDI image after subtracting the speckle estimate from the total intensity during post-processing, using \texttt{corgihowfsc} with the compact model. \textbf{Right:} Comparison of STD measurements in CDI images with the theoretical value $2I_{\mathrm{coh}}\mathrm{err}$ given by Eq.~\ref{sigma_Incoherent}  for different probe contrasts in the DH, using \texttt{corgihowfsc} compact model. The total contrast without applying the probe is $\sim 2\times 10^{-8}$. A linear fit is performed in the region of high probe contrast.}
    \label{std_incoh}
\end{figure}

During the PWP, we apply both types of probes, with a total base contrast of $2\times 10^{-8}$ in the central subband of Band 1 (see Tab. \ref{tab:cgi_modes_spectraux}), at $\lambda_c = 575$ nm and $\Delta\lambda/\lambda =3.3\%$. We plot in Fig.~\ref{std_incoh} (left) DH residuals in a CDI image for a probe contrast of $5\times 10^{-7}$. These residuals are directly caused by the nonlinear PWP errors. The values for probe voltage and total contrast correspond to the targeted operational regime for the alternative probes during the Roman Coronagraph TD1 campaign. 

When we calculate the standard deviation in CDI images for both probe sets, we observe in line with the predictions made in Fig.~\ref{err_NL} that the residuals are lower with the Gaussian probes than with the Sinc-sinc-sine probes in Fig.~\ref{std_incoh} (left). With a probe contrast of $5\times 10^{-7}$, Gaussian probes reduce the standard deviation in the CDI image by a factor of $\sim 3.5$.
We can extend the study to investigate the evolution of the standard deviation in the CDI images as a function of the probe contrast. This study is presented in Fig.~\ref{std_incoh} (right). We find, as expected, that the standard deviation of the CDI images due to nonlinear errors is proportional to the probe amplitude. However, there is a loss of agreement with the theory for smaller probe amplitudes, where a floor appears to be reached at a standard deviation of $\sim2\times10^{-10}$.

The theoretical values obtained through Eq.~\ref{sigma_Incoherent} for Gaussian probes are close to those measured in simulation, validating our model. For Sinc-sinc-sine probes, the measured error is lower than the theoretical prediction. At this stage of the study, we have not investigated this difference in detail, but it may be due to the strong differences between the modulated and unmodulated sinc probes. Equation \ref{err_prop} would then need to be reviewed, perhaps with significant weighting or covariance elements. In any case, our simulations show a decrease in the standard deviation of the CDI residuals by a factor of up to $\sim 3.5$ when switching from the Sinc-sinc-sine to the Gaussian probe in the central subband of Band 1, at $\lambda_c = 575$~nm and $\Delta\lambda/\lambda = 3.3\%$.
\section{Conclusion and future work}
\label{conclusion}
To optimize the operation of the HOWFSC loop on fainter target stars (down to $V \sim 5$), the PWP algorithm requires larger probe amplitudes to maintain a sufficient signal-to-noise ratio. However, increasing the amplitude of the nominal Sinc-sinc-sine probes inherently breaks the linear approximation of the electric field, introducing severe spatial non-uniformities.

In this study, we examined an alternative strategy for estimating the wavefront using Gaussian probes in order to mitigate these nonlinear biases. Through an analytical decomposition of the higher-order terms following Laginja et al. (2025)~\cite{laginja_extended_2025}, we show that the dominant nonlinear error depends strictly on the probe’s shape. Using the \texttt{Asterix} optical simulator, we performed a trade-off analysis to optimize the parameters of the Gaussian probes, demonstrating that a width of $\sigma_G = 1$ actuator pitch provides good illumination of the 3 to 9 $\lambda/D$ DH in Band 1 in NFOV imaging mode.

We subsequently validated this approach using \texttt{corgihowfsc}, the official Roman Coronagraph Instrument control software, coupled with its compact optical model. Our noise-free simulations confirmed that replacing the nominal probes with the optimized Gaussian probes reduces the standard deviation in the CDI images by a factor of $\sim 3.5$. This drastic reduction in the nonlinear error confirms that alternative probing is a low-risk, high-gain enhancement for the Roman Coronagraph Instrument flight software, but also a major decision regarding the use of CDI for post-processing with future space-based high-contrast telescopes.

Looking forward, several key milestones remain to fully qualify this enhanced mode for on-sky operations during the TD1 phase. Future work will focus on:
\begin{itemize}
    \item \textbf{End-to-end performance with \texttt{corgisim}~\cite{krist2023end}:} Transitioning from the noise-free compact model to the full \texttt{corgihowfsc} end-to-end simulations with \texttt{corgisim}, by taking into account other sources of noise, such as photon noise and detector noise.
    \item \textbf{Simulation of our flight test:} Checking with the official exposure time calculator and \texttt{corgisim} to confirm that the observational framework will allow us to highlight the nonlinear effects of the nominal probes compared to the Gaussian probes. 
    \item \textbf{Optical bench test (THD2):} In addition to simulations, one option would be to replicate the experiment we wish to conduct in flight on the Meudon optical testbed (THD2).
\end{itemize}

\appendix

\section{Theoretical Formalism of the PWP}
\label{annexe}
\subsection{Empirical Amplitude and Phase Extraction of the Probe}\label{annexe_1}
To reconstruct the focal plane electric field robustly, RST uses a semi-empirical Pairwise Probing (PWP) algorithm~\cite{cady_high-order_2025}. Let us introduce the coronagraphic linear operator $C$ that starts at the DM and propagates to the focal plane, representing the electric field $E_0$ as seen from the DM. The measured intensity for the unmodulated image ($I_0$) and the positive/negative modulated images ($I_m^+, I_m^-$) can be written as:
\begin{align}
    I_0 &= |C[E_0]|^2 + I_{\text{incoh}} \, , \label{eq:pwp_i0} \\
    I_m^+ &= |C[E_0e^{i\Psi_m}]|^2 + I_{\text{incoh}} \, , \label{eq:pwp_iplus} \\
    I_m^- &=  |C[E_0e^{-i\Psi_m}]|^2 + I_{\text{incoh}} \, . \label{eq:pwp_iminus}
\end{align}
 
Assuming that the amplitude of the aberrations in $E_0$ and the applied phase probes are small compared to the wavelength, we linearize the coronagraphic propagation. We define the coherent stellar speckle field $E_s = C[E_0]$ and the linear coherent probe field $P_m \equiv C[E_0\Psi_m]$. Under these assumptions, the total intensity on the science camera can be separated into three terms: $E_s$, $P_m$, and the incoherent background $I_{\text{incoh}}$ (which includes exoplanetary light, zodiacal dust, and biases). Expanding Eqs.~(\ref{eq:pwp_iplus}) and (\ref{eq:pwp_iminus}) to the first order yields:
\begin{align}
    I_m^+ &\approx |C[E_0] + iC[E_0\Psi_m]|^2 + I_{\text{incoh}} \equiv |E_s + iP_m|^2 + I_{\text{incoh}} \, , \label{eq:pwp_iplus_approx} \\
    I_m^- &\approx |C[E_0] - iC[E_0\Psi_m]|^2 + I_{\text{incoh}} \equiv |E_s - iP_m|^2 + I_{\text{incoh}} \, . \label{eq:pwp_iminus_approx}
\end{align}

From $I^+_m$ and $I^-_m$, the first order theoretical electric field of the probes $P_m$ can be isolated:
\begin{equation}
    P_m \approx \frac{C[E_0e^{i\Psi_m}]-C[E_0]}{i} \approx \frac{C[E_0]-C[E_0e^{-i\Psi_m}]}{i} \, .
\end{equation}

Because higher-order nonlinearities break this perfect symmetry, in practice, we take the average of the two approximations to cancel out the even-order phase terms and obtain a better measurement $P_{\text{effective},m}$:
\begin{equation}
   P_{\text{effective},m}= \frac{C[E_0e^{i\Psi_m}]- C[E_0e^{-i \Psi_m}]}{2i}\, .
\end{equation}

First, to obtain the probe amplitude in the focal plane, we use an empirical estimation rather than the model, demonstrating that:
\begin{align}
    \frac{I_m^+ + I_m^-}{2} - I_0 &= \frac{|C[E_0e^{i\Psi_m}]|^2 + |C[E_0e^{-i\Psi_m}]|^2}{2} - |C[E_0]|^2 \nonumber \\
    &\approx |P_m|^2 \, ,
\end{align}
which yields:
\begin{equation}
    |P_m| \approx \sqrt{\frac{I_m^+ + I_m^-}{2} - I_0}\, . \label{eq:pwp_empirical_amp}
\end{equation}

Then, we determine the model-based probe phase by taking the argument of $P_{\text{effective},m}$:
\begin{equation}
    \phi_m = \arg(P_{\text{effective}, m}) \, .
\end{equation}

The effective, semi-empirical complex probe field used for the inversion is therefore constructed as:
\begin{equation}
    P_m = |P_m| e^{i\phi_m} = \Re(P_m) + i\Im(P_m)\, .
\end{equation}

In our simulations, we use three pairs $\pm \Psi_m$ of probes $m=[1,2,3]$, providing a total of 6 probes. We assume that the 6 probes have the same amplitude: 
\begin{equation}
\label{delta_p}
    |P| = |P_m| = \sqrt{I_{\mathrm{probe}}},\, \forall m\in[1,2,3] \,.
\end{equation}
\subsection{Estimation of the Speckles Electric Field}\label{annexe_2}

To isolate the real and imaginary components of the speckle field $E_s = \Re(E_s)+i\Im(E_s)$, we subtract the modulated images. From Equations \ref{eq:pwp_iplus_approx} and \ref{eq:pwp_iminus_approx}, the intensity difference is:
\begin{align}
    I_m^+ - I_m^- &= |E_s+iP_m|^2 -|E_s-iP_m|^2 \nonumber \\
                  &= 4 \Big[ \Re(P_m)\Im(E_s) - \Im(P_m)\Re(E_s) \Big]\, .
\end{align}

We can write the linearised observation vector element $\delta_m$ as half the intensity difference:
\begin{equation}
    \delta_m = \frac{I_m^+ - I_m^-}{2} = -2\Im(P_m)\Re(E_s) + 2\Re(P_m)\Im(E_s) \, .
\end{equation}

For $m = 3$ probe pairs across a specific focal plane pixel $(k,l)$ yields the following matrix equation:
\begin{equation}
    D_{k,l} = M_{k,l}F_{k,l}\,  ,
\end{equation}
where $D$ is the probe differential vector, $M$ is the modulation matrix, and $F$ is the unknown speckle field vector we want to estimate:
\begin{equation}
    D_{k,l} = \begin{bmatrix} \delta_1 \\ \delta_2 \\ \delta_3 \end{bmatrix}_{k,l}, \quad
    M_{k,l} = \begin{bmatrix} 
        -2\Im(P_1) & 2\Re(P_1) \\
        -2\Im(P_2) & 2\Re(P_2) \\
        -2\Im(P_3) & 2\Re(P_3) 
    \end{bmatrix}_{k,l}, \quad
    F_{k,l} = \begin{bmatrix} \Re(E_s) \\ \Im(E_s) \end{bmatrix}_{k,l} \, .
\end{equation}

To estimate the electric field $E_s$, the previous matrix equation system must be inverted. We employ \texttt{numpy.linalg.lstsq}, which leverages a Singular Value Decomposition (SVD). This estimation is performed for each pixel $(k,l)$ and at each wavelength across the three observation subbands ($\lambda(\text{nm})=[546,575,604] $).
\subsection{Propagation of Nonlinear Errors and Correlation Assumption}
\label{annexe_correlation}

To derive Eq.~\ref{correlation_termes}, the expression of the PWP intensity is first recalled:
\begin{equation}
    I_{\mathrm{PWP}} = \Re(E_{\mathrm{PWP}})^2 + \Im(E_{\mathrm{PWP}})^2 \, .
\end{equation}

The statistical variance $\sigma_{I_{\mathrm{PWP,NL}}}^2$ is expressed as:
\begin{equation}
    \begin{split}
        \sigma_{I_{\mathrm{PWP,NL}}}^2 &= 4\Big[ \Re(E_{\mathrm{PWP}})^2\sigma^2_{\mathrm{NL}}(\Re(E_{\mathrm{PWP}})) + \Im(E_{\mathrm{PWP}})^2\sigma^2_{\mathrm{NL}}(\Im(E_{\mathrm{PWP}})) \Big] \\
        &\quad + 8\Re(E_{\mathrm{PWP}})\Im(E_{\mathrm{PWP}})\mathrm{Cov}(\Re(E_{\mathrm{PWP}}), \Im(E_{\mathrm{PWP}})) \, .
    \end{split}
\end{equation}

The covariance can be explicitly expressed using the correlation coefficient $\rho \in [-1, 1]$:
\begin{equation}
    \mathrm{Cov}(\Re(E_{\mathrm{PWP}}), \Im(E_{\mathrm{PWP}})) = \rho \cdot \sigma_{\mathrm{NL}}(\Re(E_{\mathrm{PWP}})) \cdot \sigma_{\mathrm{NL}}(\Im(E_{\mathrm{PWP}})) \, .
\end{equation}

For the nonlinear errors, a perfect positive correlation is assumed ($\rho = 1$). Substituting $\rho =1 $ into the variance expansion allows the entire expression to be factored as:
\begin{equation}
    \sigma_{I_{\mathrm{PWP,NL}}}^2 = \left( 2\Re(E_{\mathrm{PWP}})\sigma_{\mathrm{NL}}(\Re(E_{\mathrm{PWP}})) + 2\Im(E_{\mathrm{PWP}})\sigma_{\mathrm{NL}}(\Im(E_{\mathrm{PWP}})) \right)^2 \, .
\end{equation}

\acknowledgments 
 
This work has received support from France 2030 through the project named Académie Spatiale d'Île-de-France (\url{https://academiespatiale.fr/}) managed by the French National Research Agency (ANR) under the reference ANR-23-CMAS-0041. Furthermore, L. Delaye acknowledges financial support from SPIE through a student conference support grant to attend the 2026 Astronomical Telescopes + Instrumentation conference. This research was carried out in part at the Jet Propulsion Laboratory, California Institute of Technology, under a contract with the National Aeronautics and Space Administration (80NM0018D0004).

\bibliography{report} 
\bibliographystyle{spiebib} 

\end{document}